\documentclass[aip,jcp,preprint,amsmath,amsfonts,amssymb,a4paper,eqsecnum,superscriptaddress]{revtex4-1}

\usepackage[usenames,dvipsnames]{color}
\usepackage{color,graphicx,amsthm}
\usepackage[utf8]{inputenc}
\usepackage{natbib}

\usepackage[T1]{fontenc}

\usepackage{soul}

\begin{document}

\allowdisplaybreaks

\title{Non-adiabatic effects within a single thermally-averaged potential
energy surface: Thermal expansion and reaction rates of small molecules}

\author{J. L. Alonso}
\affiliation{Departamento de F{\'{\i}}sica Te\'orica, Universidad de Zaragoza, Pedro Cerbuna 12, E-50009 Zaragoza, Spain}
\affiliation{Instituto de Biocomputaci\'on y F{\'{\i}}sica de Sistemas Complejos (BIFI), Universidad de Zaragoza, Mariano Esquillor s/n, Edificio I+D, E-50018 Zaragoza, Spain}
\affiliation{Unidad Asociada IQFR-BIFI}
\author{A. Castro}
\affiliation{Instituto de Biocomputaci\'on y F{\'{\i}}sica de Sistemas Complejos (BIFI), Universidad de Zaragoza, Mariano Esquillor s/n, Edificio I+D, E-50018 Zaragoza, Spain}
\affiliation{Zaragoza Scientific Center for Advanced Modeling (ZCAM), Universidad de Zaragoza, Mariano Esquillor s/n, Edificio I+D, E-50018 Zaragoza, Spain}
\affiliation{Unidad Asociada IQFR-BIFI}
\affiliation {Fundaci\'on ARAID, Paseo Mar\'{\i}a Agust\'{\i}n 36, E-50004 Zaragoza, Spain.}
\author{J. Clemente-Gallardo}
\affiliation{Departamento de F{\'{\i}}sica Te\'orica, Universidad de Zaragoza, Pedro Cerbuna 12, E-50009 Zaragoza, Spain}
\affiliation{Instituto de Biocomputaci\'on y F{\'{\i}}sica de Sistemas Complejos  (BIFI), Universidad de Zaragoza, Mariano Esquillor s/n, Edificio I+D, E-50018 Zaragoza, Spain}
\affiliation{Zaragoza Scientific Center for Advanced Modeling (ZCAM), Universidad de Zaragoza, Mariano Esquillor s/n, Edificio I+D, E-50018 Zaragoza, Spain}
\affiliation{Unidad Asociada IQFR-BIFI}
\author{P. Echenique}
\affiliation{Instituto de Qu\'{\i}mica F\'{\i}sica Rocasolano, CSIC, Serrano 119, E-28006 Madrid, Spain}
\affiliation{Departamento de F{\'{\i}}sica Te\'orica, Universidad de Zaragoza, Pedro Cerbuna 12, E-50009 Zaragoza, Spain}
\affiliation{Instituto de Biocomputaci\'on y F{\'{\i}}sica de Sistemas Complejos (BIFI), Universidad de Zaragoza, Mariano Esquillor s/n, Edificio I+D, E-50018 Zaragoza, Spain}
\affiliation{Zaragoza Scientific Center for Advanced Modeling (ZCAM), Universidad de Zaragoza, Mariano Esquillor s/n, Edificio I+D, E-50018 Zaragoza, Spain}
\affiliation{Unidad Asociada IQFR-BIFI}
\author{J. J. Mazo}
\affiliation{Instituto de Ciencia de Materiales de Arag\'on, and Departamento de F\'{\i}sica de la Materia Condensada, CSIC-Universidad de Zaragoza, E-50009 Zaragoza, Spain}
\author{V. Polo}
\affiliation{Departamento de Qu{\'{\i}}mica Org\'anica y Qu{\'{\i}}mica F{\'{\i}}sica, Universidad de Zaragoza, Pedro Cerbuna 12, E-50009 Zaragoza, Spain}
\affiliation{Instituto de Biocomputaci\'on y F{\'{\i}}sica de Sistemas Complejos (BIFI), Universidad de Zaragoza, Mariano Esquillor s/n, Edificio I+D, E-50018 Zaragoza, Spain}
\affiliation{Unidad Asociada IQFR-BIFI}
\author{A. Rubio}
\affiliation{Nano-Bio Spectroscopy Group and ETSF Scientific Development Centre, Departamento de F{\'{\i}}sica de Materiales, Universidad del Pa{\'{\i}}s Vasco, E-20018 San Sebasti\'an, Spain}
\affiliation{Centro de F{\'{\i}}sica de Materiales CSIC-UPV/EHU-MPC and DIPC, E-20018 San Sebasti\'an, Spain}
\author{D. Zueco}
\affiliation {Fundaci\'on ARAID, Paseo Mar\'{\i}a Agust\'{\i}n 36, E-50004 Zaragoza, Spain.}
\affiliation{Instituto de Ciencia de Materiales de Arag\'on, and Departamento de F\'{\i}sica de la Materia Condensada, CSIC-Universidad de Zaragoza, E-50009 Zaragoza, Spain}

\begin{abstract}

At non-zero temperature and when a system has low-lying excited electronic
states, the ground-state Born--Oppenheimer approximation breaks down and the
low-lying electronic states are involved in any chemical process. In this
work, we use a temperature-dependent effective potential for the nuclei which
can accomodate the influence of an arbitrary number of electronic states in a
simple way, while at the same time producing the correct Boltzmann equibrium
distribution for the electronic part. With the help of this effective
potential, we show that thermally-activated low-lying electronic states can
have a significant effect in molecular properties for which electronic
excitations are oftentimes ignored. We study the thermal expansion of the
Manganese dimer, Mn$_2$, where we find that the average bond length
experiences a change larger than the present experimental accuracy upon the
inclusion of the excited states into the picture. We also show that, when
these states are taken into account, reaction rate constants are modified. In
particular, we study the opening of the ozone molecule, O$_3$, and show that
in this case the rate is modified as much as a 20\% with respect to the
ground-state Born--Oppenheimer prediction.

\end{abstract}

\maketitle

\section{Introduction}
\label{sec:introduction}

In the ab initio study of the behaviour of molecular systems, it is common to
perform the Born--Oppenheimer separation of the nuclear and electronic parts
of the molecular wavefunction. This approximation is based in the large
difference between the masses of the electrons and the
nuclei,\cite{Born1927,Echenique2007a} and therefore becomes exact only in the
classical limit for the nuclei. Computing the electronic energy spectrum for
different positions of the nuclei, one obtains the different so-called
Potential Energy Surfaces (PESs) ---one for the electronic ground state, and
one for each excited state. Now the question arises, how may we use these PESs
in order to produce accurate and convenient physical models? The simplest
option is ground-state Born--Oppenheimer (gsBO), or typically just
Born--Oppenheimer (BO). One may then consider the nuclei to be quantum
particles and solve their corresponding Schr{\"{o}}dinger equation, or take
the classical limit and perform BO Molecular Dynamics. In any case, a model
based on only one PES (usually the ground state one) is an \emph{adiabatic}
approximation, based on the neglect of the non-adiabatic couplings.

However, at non-zero temperature and when a system has low-lying excited
electronic states, these electronic states are involved in any chemical
process, and their influence produces the so-called non-adiabatic effects. In
this paper, we use a thermodynamically accurate generalization, introduced in
Ref.~\onlinecite{Alonso2010}, of the gsBO potential, built with the PESs of
the lowest-lying electronic states appropriately weighted to produce the
correct Boltzmann equilibrium distribution of the electronic part, to
specifically study the thermal expansion and reaction rates of small
molecules. For both phenomena, a substantial amount or research is constructed
on the implicit assumption of \emph{one} electronic potential, which may be
fitted to experimental results or computed ab initio. The possibility of
electronic excitations is typically either ignored or handled by independently
considering each of the excited PES (although we can mention
Ref.~\onlinecite{Mazzola2012} as a very recent exception to this general
trend, with similar aims to the ones we pursue here). The thermally averaged
potential energy surface that we use in this work permits to include
electronic excitations while still preserving the single potential
methodology.

The possibility, in which our formalism is based, of dividing a system of
particles into a quantum and a classical subsystem (typically, electrons and
nuclei) is of wide interest in several areas of physics and chemistry. If the
temperature is of the order of the electronic gap or larger and excited
electronic energy levels have to be included in the formalism, a variety of
approaches can be considered.\cite{Zhu2005, Truhlar2007, schmidt:044104,
Worth2004, Martinez1996, Nakamura2002, Baer2006, Prez1999} For example,
the decay-of-mixing method by Truhlar and coworkers \cite{Zhu2005,Truhlar2007}
constitutes a powerful mixed quantum-classical scheme for modeling
non-Born-Oppenheimer chemistry, although the incorporation of temperature to
these methods has not been studied as far as we are aware. In the case of
Ehrenfest dynamics, which also includes non-adiabatic effects at a different
level, the temperature has been introduced through the formulation of
Nos\'e.\cite{Alonso2011, [{It is also worth remarking that some common criticisms
that stated that Ehrenfest is a fully coherent method, and thus it cannot account
for such important effects as mixing and decoherence, have recently been
challenged: }] Alonso2012}

The temperature dependence of molecular properties (geometry and vibrational
frequencies) of free molecules has been a subject of research for more than 60
years,\cite{Bartell1955} with new theoretical analyses coming out still in
very recent times.\cite{Varga2010,Varga2011} From the experimental viewpoint,
on the other hand, several studies of hot molecules
\citep{Bartell1979,Bartell1979a,Goates1982,Goates1982a,Bartell1984} have found
thermal expansion of bond lengths. In this paper we calculate the temperature
dependence of the bond length in the case of diatomic molecules in which
nuclei can be treated as classical particles. Although bond-length expansion
as the temperature increases is expected even for temperature-independent
potentials such as the gsBO one, the use of our thermodynamically more
accurate effective potential adds new non-adiabatic effects which, as we will
show for the Mn$_2$ dimer, may significantly alter the final quantitative
results if low-lying excited states are present. These effects must be
considered when performing the necessary rovibrational averaging in order to
compare accurate theoretical and experimental results.

The temperature dependence of the transition rate is also a constant subject
of study \cite{Hanggi1990,Fleming2011,Bothma2010} and, more recently, the
question has been asked of whether or not quantum tunneling below the energy
barrier significantly enhances the reaction
rate.\cite{Fleming2011,Bothma2010,Iyengar2008} Using a general framework to
describe tunneling,\cite{Bothma2010} it is shown that tunneling below the
barrier only occurs for temperatures less than a reference one, denoted by
$T_0$, and which is determined by the curvature of the ground-state PES
(gsPES) at the top of the barrier. However, at non-zero temperatures and when
a system has low-lying excited electronic states, all estimates based on the
ground-state PES should be reconsidered.

As demonstrated in this work using our thermodynamically accurate
generalization of the gsBO potential, the inclusion of low-lying electronic
states into the picture may significantly alter the reaction rates and the
curvature near the top of the barrier. As a model system to exemplify the
added effects, we have chosen the transition between the open and closed forms
of the ozone molecule, in which the barrier lies in a region where there is an
avoided crossing between the ground and first-excited electronic states.
Having no problems whatsoever with avoided-crossing situations, our effective
potential is a convenient choice to account for the influence of low-lying
excited states in this phenomenon at non-zero temperature.

In Sec.~\ref{sec:methods}, we provide a comprehensive summary of the
definition and meaning of the effective potential introduced in
Ref.~\onlinecite{Alonso2010} which includes the effect of excited electronic
states and which will be used throughout the document. In
Sec.~\ref{subsec:thermal_expansion}, we present the first application of the
effective-potential technique: the study of the thermal expansion of the
Manganese dimer, Mn$_2$; in Sec.~\ref{subsec:reaction_rate}, we discuss the
influence of the inclusion of low-lying excited electronic states on the
reaction rate of small molecules. Reaction rates are affected on the one hand
by the energy barrier reduction. On the other hand, the curvature at the top
of the barrier is smaller for our effective potential than for the gsPES. In
this section we will also study the case of the opening of the ozone molecule,
O$_3$. Finally, in Sec.~\ref{sec:conclusions}, we comment on the most
important conclusions of this work and highlight some possible implications
and future lines of research.

\section{Theoretical framework}
\label{sec:methods}

Let us have a quantum-classical system formed by $N$ classical particles
described by their Euclidean coordinates $R := (\vec{R}_1,\ldots,\vec{R}_N)$
and momenta $P := (\vec{P}_1,\ldots,\vec{P}_N)$, and $n$ quantum particles
described by an $n$-body wavefunction $|\psi\rangle$. The starting point of
our formalism is the assumption that the following is an accurate formula to
compute canonical equilibrium expected values of quantum-classical observables
$\hat{O}(R,P)$:
\begin{equation}
\label{eq:expO_RP}
\langle \hat{O}(R,P) \rangle =
 \frac{\displaystyle \int dRdP \, \mathrm{tr} \left( \hat{O}(R,P) \,
       e^{- \frac{\hat{H}(R,P)}{k_{\rm B}T}} \right)}
      {\displaystyle \int dRdP \, \mathrm{tr} \left(
       e^{- \frac{\hat{H}(R,P)}{k_{\rm B}T}} \right)} \,,
\end{equation}
where $k_{\rm B}$ is the Boltzmann constant and $T$ the temperature, $dRdP$
denotes integration over all position and momenta in the appropriate ranges.
The object $\hat{H}(R,P)$ is the quantum-classical Hamiltonian, which, in the
case of molecular systems, takes the form
\begin{equation}
\label{eq:H}
\hat{H}(R,P) := \hat{1} \sum_{K=1}^N \frac{\vec{P}_K^2}{2 M_K}
   + \hat{H}_e(R) \,,
\end{equation}
where $\hat{1}$ denotes the identity matrix, $M_K$ is the mass of the $K$-th
nucleus, and the \emph{electronic Hamiltonian}, $\hat{H}_e(R)$,
contains all particle interactions and the electronic kinetic term (see
Ref.~\onlinecite{Echenique2007a} for an explicit expression).

It is also convenient to write Eq.~(\ref{eq:expO_RP}) as
\begin{equation}
\label{eq:expO_RP_2}
\langle \hat{O}(R,P) \rangle =
 \int dRdP \, \mathrm{tr} \left( \hat{O}(R,P) 
  \hat{\rho}_\mathrm{eq}(R,P) \right) \,,
\end{equation}
in terms of a $(R,P)$-dependent equilibrium density matrix, defined by
\begin{equation}
\label{eq:rhoeq_RP}
\hat{\rho}_\mathrm{eq}(R,P) := 
\frac{\displaystyle e^{-\frac{\hat{H}(R,P)}{k_{\rm B}T}}}
     {\displaystyle \int dR^\prime dP^\prime \, \mathrm{tr} \left(
      e^{-\frac{\hat{H}(R^\prime,P^\prime)}{k_{\rm B}T}} \right)} \,.
\end{equation}

As shown in Ref.~\onlinecite{Alonso2010}, the justification of
Eq.~(\ref{eq:expO_RP}) for computing equilibrium expected values in
quantum-classical models stems on the one hand from plausibility arguments
related to the classical limit of the behaviour of the nuclei, such as the
ones given in Ref.~\onlinecite{Mauri1993}. However, it can also be obtained as
the zero-th order approximation (in the quantum-classical mass ratio) to the
canonical equilibrium associated with an, in principle, more rigorous
quantum-classical formulation based on the Wigner formalism, as shown by
Kapral and Ciccotti \cite{Kapral1999} and Nielsen et al. \cite{Nielsen2001}.
It may also be rationalized by entropic arguments\cite{entropy}.
In any case, and as it can be seen in the references in
Ref.~\onlinecite{Alonso2010}, irrespective of how good an approximation
Eq.~(\ref{eq:expO_RP}) is for any given application ---always a difficult
question---, it is often the desired, target expectation value when designing
quantum-classical schemes.

The main realization in which the effective potential that we will use in this
work is based is that, for observables which do not depend explicitly on the
electronic degrees of freedom, i.e., which are of the form
\begin{equation}
\label{eq:O_classical}
\hat{O}(R,P) = \hat{1} \, O(R,P) \,,
\end{equation}
where $O(R,P)$ is a number, the target expected value in
Eq.~(\ref{eq:expO_RP}) can be rewritten as
\begin{equation}
\label{eq:expO_RP_eff}
\langle O(R,P) \rangle =
 \frac{\displaystyle \int dRdP \, O(R,P) \,
                     e^{-\frac{H_\mathrm{eff}(R,P;T)}{k_{\rm B}T}} }
      {\displaystyle \int dRdP \,
                     e^{-\frac{H_\mathrm{eff}(R,P;T)}{k_{\rm B}T}} } \,.
\end{equation}

Now, $H_\mathrm{eff}(R,P;T)$ is a purely classical, $T$-dependent, effective
Hamiltonian defined as
\begin{eqnarray}
\label{eq:Heff}
H_\mathrm{eff}(R,P;T) & := & - k_{\rm B}T \, \mathrm{ln}
  \, \mathrm{tr} \, e^{-\frac{\hat{H}(R,P)}{k_{\rm B}T}} \nonumber \\
 & = & \sum_{K=1}^N \frac{\vec{P}_K^2}{2 M_K}
  - k_{\rm B}T \, \mathrm{ln}
  \, \mathrm{tr} \, e^{-\frac{\hat{H}_e(R)}{k_{\rm B}T}} \nonumber \\
& =: & \sum_{K=1}^N \frac{\vec{P}_K^2}{2 M_K} + V_\mathrm{eff}(R;T) \,,
\end{eqnarray}
where we have used Eq.~(\ref{eq:H}). In the last line, we have finally defined
the promised, purely classical, $T$-dependent, $P$-independent, effective
potential
\begin{equation}
\label{eq:Veff}
V_\mathrm{eff}(R;T) := - k_{\rm B}T \, \mathrm{ln}
  \, \mathrm{tr} \, e^{-\frac{\hat{H}_e(R)}{k_{\rm B}T}} \,,
\end{equation}
which can be used to describe the behaviour of the nuclei in a classical
mechanical setting, producing the correct target equilibrium in
Eq.~(\ref{eq:expO_RP}) for classical observables ---by construction---, and
including the influence of all the electronic excited states.

Indeed, if we consider the adiabatic basis $\{|\psi_k(R)\rangle\}$, which
diagonalizes the electronic Hamiltonian $\hat{H}_e(R)$ for each fixed position
of the nuclei,
\begin{equation}
\label{eq:adiabatic_basis}
\hat{H}_e(R) |\psi_k(R)\rangle = E_k(R) |\psi_k(R)\rangle \,,
\end{equation}
being $\{E_k(R)\}$ the corresponding PESs, i.e., the eigenvalues of the
electronic Hamiltonian as a function of the nuclear positions, then we can
rewrite the effective potential in Eq.~(\ref{eq:Veff}) as
\begin{eqnarray}
\label{eq:Veff_basis}
V_\mathrm{eff}(R;T) & = & - k_{\rm B}T \, \mathrm{ln}
 \sum_k e^{-\frac{E_k(R)}{k_{\rm B}T}} \\
 & = & E_0(R) - k_{\rm B}T \, \mathrm{ln} \left[ 1 + 
  \sum_{k > 0} e^{-\frac{\Delta E_{k0}(R)}{k_{\rm B}T}} \right] \,, 
  \nonumber
\end{eqnarray}
where we have defined
\begin{equation}
\label{eq:DeltaE}
\Delta E_{k0}(R) := E_k(R) - E_0(R) \,.
\end{equation}

The expression in Eq.~(\ref{eq:Veff_basis}) explicitly shows the difference
between the ground state PES, $E_0(R)$, i.e., the gsBO potential, and the
effective potential $V_\mathrm{eff}(R;T)$ introduced in
Ref.~\onlinecite{Alonso2010} and used in this work. In particular, it is
worth remarking that
\begin{itemize}
\item At $T=0$, our effective potential becomes the gsBO one. Indeed, it is 
easy to see from Eq.~(\ref{eq:Veff_basis}) that
\begin{equation}
\label{eq:limT0}
\lim_{T \to 0} V_\mathrm{eff}(R;T) = E_0(R) \ , \quad \forall R \,.
\end{equation}

\item The same reasons that produce the previous result allow us to include
in the sum in Eq.~(\ref{eq:Veff_basis}) only the lowest-lying electronic
states and still get a good enough approximation to the exact expression if
the temperature is low compared to the states we are neglecting, i.e., if
$k_{\rm B}T \ll \Delta E_{k0}(R)$ for them. This fact will be used in the
practical cases presented in the next sections.

\item It can be seen from Eq.~(\ref{eq:Veff_basis}) that an exact
property of the effective potential is that it is strictly lower than the
gsPES, i.e., $V_\mathrm{eff}(R;T) \leq E_0(R)$, $\forall R,T$. However, since
an additive constant in a potential energy is not measurable, it must be
noticed that this inequality is relevant only inasmuch the difference $E_0(R)
- V_\mathrm{eff}(R;T)$ actually depends on $R$. See
Secs.~\ref{subsec:thermal_expansion} and~\ref{subsec:reaction_rate} for
concrete examples of this situation.

\item As we discussed in Ref.~\onlinecite{Alonso2010}, from
Eq.~(\ref{eq:Veff_basis}), we see that, if the second derivative of $\Delta
E_{10}$ at a barrier top $q_B$ verifies
\begin{equation}
\label{eq:der2_DE10}
\frac{\partial^2 \Delta E_{10}}{\partial q^2}(q_B)
 \left( 1 + e^{\frac{\Delta E_{10}(q_B)}{k_{\rm B}T} } \right) >
  \frac{1}{k_{\rm B}T}
 \left( \frac{\partial \Delta E_{10}}{\partial q}(q_B) \right)^2 \,,
\end{equation}
then we can prove
\begin{equation}
\label{eq:der2_Veff}
\left| \frac{\partial^2 V_{\rm eff}}{\partial q^2}(q_B;T) \right| <
\left| \frac{\partial^2 E_0}{\partial q^2}(q_B) \right| \,,
\end{equation}
i.e., the curvature of the effective potential at the barrier top is smaller
than the one associated to the gsPES. In avoided crossings ---a very
interesting general case, and the one actually studied in
Sec.~\ref{subsec:reaction_rate}--- since the first excited state approaches
the gsPES and then recedes from it, we have that the derivative $(\partial
\Delta E_{10} / \partial q)(q_B)$ will be approximately zero and the condition
in Eq.~(\ref{eq:der2_DE10}) will be approximately satisfied, together with
Eq.~(\ref{eq:der2_Veff}).
\end{itemize}

  Note that in order to obtain the effective
  potential, and the corresponding averages, it is not necessary to
  compute the non-adiabatic couplings. This is a consequence of the
  fact that the purpose of the effective potential is the computation
  of averages at canonical thermal equilibrium, and not the dynamics
  that may lead to it. The non-adiabatic couplings are necessary to
  carry the system to equilibrium -- by providing the necessary
  channels to couple the various electronic states. Because of this,
  we are considering that the difference between these averages and
  the ones that would be obtained by considering the gsPES only is a
  \emph{non-adiabatic effect}. However, the equilibrium averages
  predicted by Eq.~(\ref{eq:expO_RP}) can be obtained without explicit
  consideration of the couplings. The magnitude of those couplings
  might be very relevant to compute the speed of the thermalization:
  small couplings may necessitate long thermalization times, but those
  analysis are beyond the scope of the present work.  

  In this respect, also note that \emph{all} excited
  electronic states are included in the definition of the effective
  potential -- because all states are included in Boltzmann's
  equilibrium formula, regardless of how they may be coupled by
  external fields or, in the quantum-classical case, by the
  non-adiabatic couplings. The weight of each state is entirely
  determined by its energy. Any state is present in the averaging,
  even if its couplings to the ground state and to any other
  \emph{accessible} state is zero. This is an example of ergodic
  difficulties, and obviously, a dynamical averaging would not include
  such a state unless it is already included in the initial state
  sampling.  

  The straightforward application of our scheme would
  then be inadequate if one is interested in computing a “restricted
  equilibrium” average, in which a state (or full subspace of states)
  is known to be absent, due to symmetry rules. The ``experimental
  average'' would not contain those states, even if the true canonical
  ensemble average does. However in this situation it would be easy to
  correct our scheme by simply not including the forbidden states in
  the formulas.

\section{Results}
\label{sec:results}

\subsection{Non-adiabatic effects on the thermal expansion of the Mn$_2$ dimer}
\label{subsec:thermal_expansion}

Theoreticians usually identify the ``molecular structure'' with the
\emph{equilibrium structure}, i.e., the point determined by the absolute
minimum of the ground state Born-Oppenheimer potential energy surface (gsPES).
This point in $\mathbb{R}^{3N}$ space ($N$ being the number of atoms) is well
defined and has an easy intuitive meaning: the position occupied by the nuclei
at equilibrium, if they had infinite mass (in which case they would be
classical point particles). This geometry corresponds to a motionless
molecule, that does not exist because molecules vibrate and move even at zero
temperature. Therefore, this equilibrium structure is a theoretical concept
that is not provided ---at least not directly--- by the experimental
techniques utilized to probe molecular structure.

In fact, different experimental techniques yield different \emph{averaged}
results, whose value and meaning depend on the physical process involved in
the measurement. For example, X-ray diffraction provides distances
between the electronic charge distribution centroids. Gas-phase electron
diffraction (ED), on the other hand, provides internuclear distances.
Microwave spectroscopy measures moments of inertia, which can be directly
related to nuclear distances to the molecular center of mass. Infrared, Raman,
and ultraviolet spectroscopies can also be used for complementary analysis. In
all cases, the results are averaged over the populated rotational and
vibrational molecular states ---and, as we will show below, possibly over
different electronic potential energy surfaces. Those techniques achieve a
remarkable precision (of the order of 0.001 \AA), but nevertheless provide
different numbers each one of them.

\begin{figure}[b]
\centerline{\includegraphics[width=6cm,angle=-90]{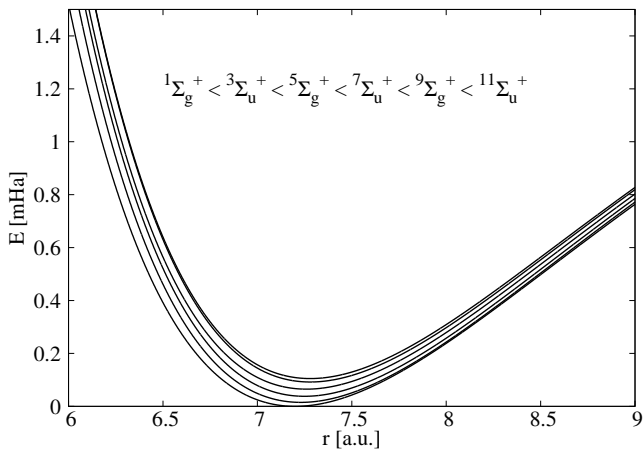}}
\caption{
\label{fig:Mn2pes}
Six lowest lying PESs of the Mn$_2$ dimer, taken from
Ref.~[\onlinecite{tzeli:154310}]. The calculations were performed through
multireference variational calculations coupled with augmented quadruple
correlation consistent basis sets. These PESs correspond to the ground state
``manifold'', that correlates to ground state Mn atoms.
}
\end{figure}

In order to compare the results obtained in different experiments and in
precise ab initio calculations, it is necessary to use a ``common
denominator'' representation, which can very well be the equilibrium structure
mentioned above, usually called the \emph{r$_e$ structure}. One must therefore
know how to relate the experimental result to this concept. In an ED
experiment at a given temperature, for example, one obtains the so-called
\emph{r$_a$ structure}, an operational concept with no clear physical meaning.
It can be related, however, to the thermally averaged internuclear distance,
or \emph{r$_g$ structure}, which is of physical significance. It is not
equivalent to the $r_e$ structure, not even at 0 K, because of the vibrational
and rotational (also called \emph{centrifugal}) distortions. The relationship
between $r_e$ and an averaged structure such as $r_g$ is not straightforward,
but must be considered if we want to validate high precision theoretical ab
initio calculations with experimental results or vice versa. This relationship
was first considered by Bartell,\cite{bartell:1219} and has later been
developed by several authors.\cite{kuchitsu:2460, kuchitsu:1945, Toyama1964193, herschbach:1668, springerlink:10.1023/A:1011908303949, Butayev1986}
It was soon realized that, in general, molecules \emph{expand} as temperature
increases, due to both the anharmonicity of the vibrations and to the
centrifugal ``force'' that rotations exert on the structure.

In all these studies, an assumption is implicitly made: the electrons are
adiabatically tied to their ground state, and therefore the analysis is
performed by considering the gsPES only. However, as discussed in the
introduction, the existence of non-zero non-adiabatic couplings permits the
system to visit electronic excited states, and the thermodynamic averaging
should account for this possibility if the energy gap between the gsPES and
the excited ones is not large in comparison with the thermal energy, $k_{\rm
B} T$. If calculations and experiments are to be succesfully compared, it is
therefore necessary to consider the possible effect of the excited electronic
states. We propose the use of our thermally averaged potential energy surface
introduced in the previous section for this purpose.

In principle, the rotational-vibrational averaging necessary to compute $r_g$
must be performed assuming quantum-mechanical nuclei. This is obvious at 0 K,
where the zero-point vibration would be completely absent in a classical
treatment. However, we are mostly interested in the high-temperature regime
---where the excited electronic states are expected to play a larger role and
the system becomes classical. This fact can be exemplified by looking at
closed ---albeit approximate--- theoretical formulae that exist for the
simplest cases on which we concentrate in this study: diatomic molecules. By
truncating the Taylor expansion of the PES around the equilibrium bond length,
i.e.:
\begin{equation}
\label{eq:vr}
V(r) = \frac{1}{2}k_2 (r-r_e)^2 - k_3(r-r_e)^3\,,
\end{equation}
Toyama et al.~\cite{Toyama1964193} computed, in the quantum case, the
temperature-induced variation of the average internuclear distance (with
respect to $r_e$) as:
\begin{eqnarray}
\label{eq:toyama}
\langle \Delta r \rangle (T) & := & \langle r \rangle (T) - r_e \nonumber \\
  & = & \frac{2k_{\rm B} T}{k_2 r_e} + \frac{3k_3\omega_e}{2k_2^2}
\coth\left[ \frac{\omega_e}{2k_{\rm B} T} \right]
\,,
\end{eqnarray}
where $\omega_e$ is the vibrational frequency at equilibrium ($\omega_e =
\sqrt{k_2/\mu}$, where $\mu$ is the reduced mass of the two nuclei).

The first term in Eq.~(\ref{eq:toyama}) is the \emph{centrifugal distortion};
it arises from the harmonic potential term in Eq.~(\ref{eq:vr}), and can be
considered as the variation in the average bond length caused by the overall
molecular rotations. Interestingly, a classical treatment leads to the same
expression for this term. The second term, on the contrary, has a genuinely
quantum behaviour at low temperatures. In fact, it does not approach zero as
$T\to 0$ K, producing a zero-point vibration variation of the bond length with
respect to $r_e$. We can now take the classical limit by considering $\mu \to
\infty$, in which case this second term becomes linear in $T$: $3k_3k_{\rm B}
T/k_2^2$. But note that the same behaviour occurs if we take $k_{\rm B} T$
large with respect to $\omega_e$ ---i.e., the system becomes classical for
large enough temperatures.

In view of this, and since we are interested in the high-temperature regime in
which our effective potential may significantly differ from the gsPES, we have
used the classical approximation to compute the average bond length. To this
end, we used Eq.~(\ref{eq:expO_RP_eff}), which for the dimer case (and
considering that the function $O(R,P)$ is in this case nothing else than the
internuclear distance $r = \vert\vec{R}_1 - \vec{R}_2\vert$), reads:
\begin{equation}
\label{eq:rTeff}
\langle r \rangle_{\rm eff} (T) = \frac{\int_0^L\!{\rm d}r\; r^3 e^{-\frac{V_{\rm eff}(r;T)}{k_{\rm B}T}}}
{\int_0^L\!{\rm d}r\; r^2 e^{-\frac{V_{\rm eff}(r;T)}{k_{\rm B}T}}}\,.
\end{equation}
Note the presence of a somewhat arbitrary upper limit of integration $L$.
This value cannot be made arbitrarily large, since in the limit $L\to\infty$,
the equilibrium bond length is also infinite (assuming that, as it is always
the case, the potential function remains finite at large internuclear
distances). In fact, at equilibrium and at any non-zero temperature, a dilute
gas of diatomic molecules in an infinite space does not really contain dimers
but isolated atoms. In the real world, dimers exist because there is always
some form of container, or they are in a very long-lived metastable state. In
practice, one must choose a value of $L$ such that the results are almost
insensitive to small variations of it ---acknowledging that, if $L$ is
increased to very large values, the value of $\langle r\rangle (T)$ will start
growing.

The question that we want to answer here is whether or not the use of the effective
potential in Eq.~(\ref{eq:rTeff}) leads to significantly different results
with respect to the results obtained using only the gsPES, i.e.:
\begin{equation}
\label{eq:rT0}
\langle r \rangle_0 (T) = \frac{\int_0^L\!{\rm d}r\; r^3 e^{-\frac{E_0(r)}{k_{\rm B}T}}}
{\int_0^L\!{\rm d}r\; r^2 e^{-\frac{E_0(r)}{k_{\rm B}T}}}\,.
\end{equation}
The answer cannot of course be universal, and depends on the chosen system and
the temperature regime observed. In order to illustrate the issue, we have
concentrated on the Mn$_2$ molecule; a Van der Waals weakly-bound molecule and
a specially difficult theoretical
case,\cite{Wang2004395,yamamoto:124302,tzeli:154310} for which a good feedback
between experiments and theory could help validate the conflicting theoretical
results. As we shall show, the existence of very low-lying excited electronic
states in the vicinity of the equilibrium distance has an important effect on
$\langle r \rangle(T)$, and therefore it is crucial to consider them to make
proper comparisons.

\begin{figure}
\centerline{\includegraphics[width=6cm,angle=-90]{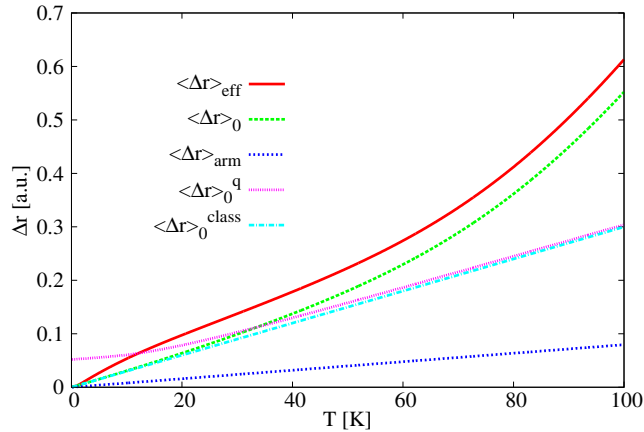}}
\caption{
\label{fig:rT}
Calculated variation of the average internuclear distance as a function of temperature for Mn$_2$.
$\langle \Delta r \rangle_{\rm eff}$ is the result computed using, in Eq.~\ref{eq:rTeff},
the effective potential including all six lowest-lying states in
Fig.~\ref{fig:Mn2pes}, while $\langle \Delta r \rangle_0$ is only computed with
the gsPES. $\langle \Delta r \rangle_0^q$ is the third-order approximation to
the quantum result, as calculated using Eq.~(\ref{eq:toyama}), and $\langle
\Delta r\rangle_0^{\rm class}$ is its classical limit. $\langle \Delta r
\rangle_{\rm arm}$ is the centrifugal term, common to all cases.
}
\end{figure}

To build our averaged potential defined in Eq.~\ref{eq:Veff_basis}, we depart from the potential energy curves provided by Tzeli et
al.~\cite{tzeli:154310}, computed from first principles with very accurate
multireference variational calculations. We only consider the six
almost-degenerate, lowest-lying states, displayed in Fig.~\ref{fig:Mn2pes}. We
have adjusted these curves to Morse functions, i.e.:
\begin{equation}
V(r) = D\left[
e^{-2\alpha\frac{r-r_e}{r_e}} - 2e^{-\alpha\frac{r-r_e}{r_e}} 
\right] + V_\infty\,.
\end{equation}
By finding the best match for the parameters $D,\alpha,r_e$ and $V_\infty$, an
almost perfect fit can be obtained with respect to the results of Tzeli et al.
The plot clearly shows the reasons for choosing Mn$_2$ in this study: the
lowest-lying states are extremely close to each other, and the potential well
is very shallow.

The results are shown in Fig.~\ref{fig:rT}. The two key curves are the ones
denoted by $\langle\Delta r\rangle_{\rm eff}$ and $\langle \Delta r\rangle_0$,
which are the results obtained with Eqs.~(\ref{eq:rTeff}) and (\ref{eq:rT0}),
respectively. The difference due to the use of the effective potential instead
of the ground state one is notable. It is larger than the resolution of modern
experimental techniques, even in the lower-temperature range. One may then
conclude that any assessment of the quality of a theoretical model based on a
comparison to experimental results should consider the influence of these
low-lying electronic excited states.

In Fig.~\ref{fig:rT}, we also display the approximate quantum result given by
Eq.~(\ref{eq:toyama}), and denoted by $\langle \Delta r \rangle_0^q$. This
quantum curve is only valid at low temperatures, since it stems from a
perturbative truncation of the potential. We display it in order to
demonstrate how the system behaves almost classically in most of the
temperature range of the plot, thus justifying our classical treatment.
Indeed, if we plot the classical limit ($\mu \to \infty$) of
Eq.~(\ref{eq:toyama}), denoted by $\langle \Delta r\rangle_0^{\rm class}$, it
quickly becomes almost identical to $\langle \Delta r \rangle_0^q$. In this
temperature region, our classical calculation, which does not truncate the
potential curve, should be almost exact. Finally, for completeness, the curve
$\langle \Delta r \rangle_{\rm arm}$ is the centrifugal term ---the difference
with the rest of the curves would be the vibrational contribution in each
case.

\begin{figure}
\centerline{\includegraphics[width=6cm,angle=-90]{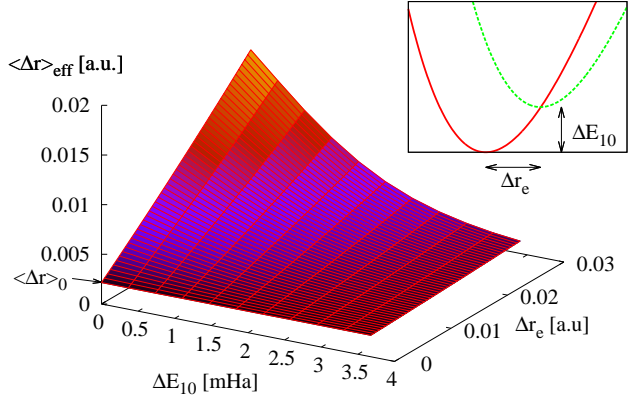}}
\caption{
\label{fig:rT2D}
Variation of the average internuclear distance, at 300 K, for a fictitious
dimer with two relevant PES, as schematically shown in the inset. The results
are given as a function of the displacement of the excited PES with respect to
the gsPES, in two directions: variation of the energy minimum $\Delta E_{10}$,
and difference in the position of this minimum, $\Delta r_e$. }
\end{figure}

Beyond this particular example, a more general question has to be addressed
when must one expect the electronic excited states to influence the thermally
averaged internuclear distances ---and therefore any experimental measurement
of molecular geometry? A simple visual inspection of a few lowest lying
excited PES, and a simple calculation with our thermally averaged PES should
give us a quick answer. Two key parameters should be carefully examined: the
``gap'', or difference between the gsPES and the closest excited ones, and how
much the position of the minima of the two curves differ. This fact is
illustrated in Fig.~\ref{fig:rT2D}, where we have considered a fictitious
dimer with two closely lying PES. The parameters of the gsPES correspond to
the Hydrogen molecule, whereas the first excited PES is a rigid displacement
in two directions: varying the minimum energy ($\Delta E_{10}$), and the
position of the corresponding minimum ($\Delta r_e$).

The 2D plot displays in Fig.~\ref{fig:rT2D} $\langle \Delta r\rangle_{\rm
eff}$ at room temperature (300 K). As the gap becomes large, the results
converge towards $\langle \Delta r\rangle_{0}$, the thermal expansion entirely
due to the gsPES. The plot shows how, for the results to differ significantly,
the gap should not be larger than a few mHa ---which is easily understood
since, at 300 K, $k_{\rm B}T$ is approximately 1 mHa. But also note that, even
if the gap is small, there is no change with respect to $\langle \Delta
r\rangle_{0}$ if the positions of the minimum of the two curves do not differ
($\Delta r_e$ is small). In other words, if the two PES are merely a rigid
vertical shift of each other, the thermally averaged PES is also a rigid
vertical shift and nothing change.

\subsection{Non-adiabatic effects on the reaction rate and the opening of ozone }
\label{subsec:reaction_rate}

\begin{figure}[b]
\centering{\includegraphics[angle=0,width=8.cm]{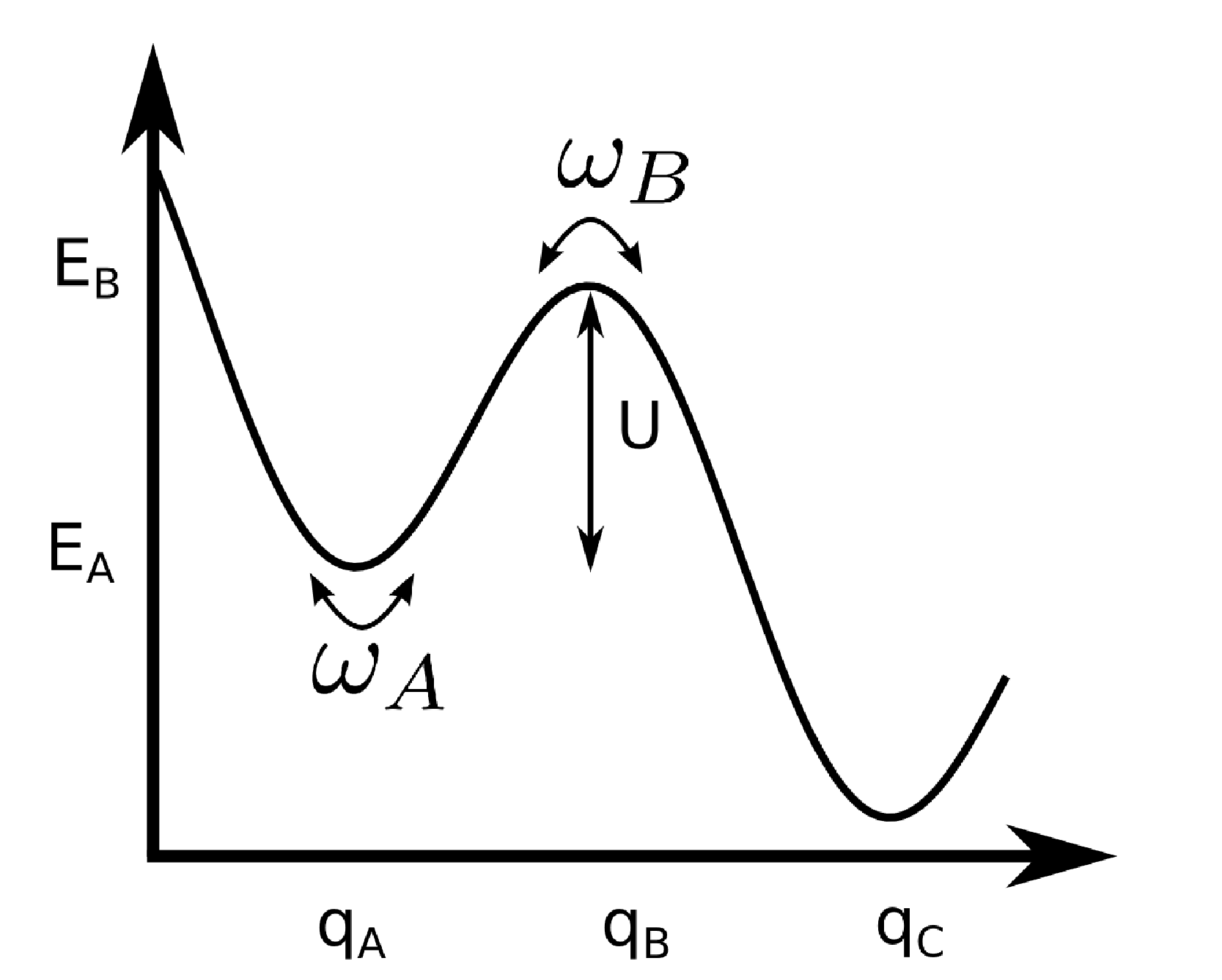}}
\caption{Sketch for the usual reaction rate problem, in terms of a generic reaction coordinate $q$: Two metastable states at
$q_A$ and $q_C$ are connected by a barrier with height $U=E_B-E_A$ and maximum
at $q_B$. The curvatures are also shown.}
\label{fig:sketch}
\end{figure}

Reaction-rate theory focuses on studying the behavior at long times of systems
with different equilibrium states. This is a subject of great interest in many
biological, chemical and physical problems. As noted by Arrhenius in 1889,
\cite{Arrhenius1889} the cornerstone of the theory is the $e^{-A/T}$
temperature dependence of the reaction rates. Such dependence can be
understood in the framework of a transition-state theory where the system
evolves as a function of a given reaction coordinate $q$ from the metastable
state $A$ to the metastable state $C$ through an energy barrier $B$, being the
activation constant related with the barrier energy $U=E_B-E_A$ (see
Fig.~\ref{fig:sketch}).

In general the reaction rate can be written as
\begin{equation}
r = k\, e^{-U/k_{\rm B}T} \,,
\label{rr}
\end{equation}
where the prefactor $k$ depends on the temperature $T$, the friction
coefficient or `damping' of the system, and the details of the potential
energy function. An accurate estimation of this prefactor has been shown to be
a formidable task and many articles have been devoted to this end ---deserving
an special mention the celebrated one by Kramers \cite{Kramers1940}. See also
Refs.\onlinecite{Pollak2005, Hanggi1990, Melnikov1991}.

As it has been shown in Ref.~\onlinecite{Alonso2010} and summarized in the
previous sections, the inclusion of excited electronic levels becomes
important at certain temperatures for obtaining suitable PESs of molecular
systems, different from the gsBO one. The object of this section is to
consider the effect of these excited states on the thermal-activation rate
calculations.

Let us denote by $r_{\rm eff}$ the activation rate for the effective potential
$V_{\rm eff}(q;T)$ in Eq.~(\ref{eq:Veff_basis}), and by $r_0$ the rate for the
gsPES, $E_0(q)$ (both of them expressed as a function of the reaction coordinate
$q$). The main effect of the inclusion of the new terms associated to the excited
electronic states is a reduction of the energy barrier from
$U_0=E_0(q_B)-E_0(q_A)$ to $U_{\rm eff}=V_{\rm eff}(q_B^{\rm eff})-V_{\rm
eff}(q_A^{\rm eff})$. We will have in general $q_A^{\rm eff} \simeq q_A$ and
$V_{\rm eff}(q_A^{\rm eff}) \simeq E_0(q_A)$, and thus the change in the energy
barrier is given by the change at the potential maximum. Regarding the activation
rate, in the simplest picture, we find that $r_{\rm eff} \sim e^{-U_{\rm
eff}/k_BT}$ and $r_{\rm eff}/r_0 \sim e^{-(U_{\rm eff}-U_0)/k_{\rm B}T}$.
Therefore, if $U_{\rm eff}-U_0$ is large enough, the effect of the energy barrier
reduction on the activation rate will be important. In addition to this effect,
it is also important to realize than the activation rate will also show a
deviation from the expected $e^{-U/k_{\rm B} T}$ temperature dependence due to
the fact that $U_{\rm eff}$ itself depends on $T$. With this in mind, the
importance of the excited electronic states can be studied looking for deviations
of $r(T)$ from its expected dependence.

A more detailed analysis must take into account the influence of the prefactor
$k$ too. To this end, some approximations need to be done. Let us place our
problem in the context of the large-barrier ($U/k_{\rm B}T \gg 1$) and
strong-friction limit. In such a situation, we learnt from Kramers that,\cite{Kramers1940}
\begin{equation}
  r_{\rm KHD}=\frac{\omega_A}{\gamma} 
  \sqrt{\frac{k_{\rm B} T}{2\pi}} 
  \left \{ \int_{q_A}^{q_B} {\rm e}^{U(q)/k_{\rm B}T} dq \right \}^{-1} \,,
\label{I}
\end{equation}
where $U(q) := V(q) - V(q_A)$, being $V(q)$ the appropriate potential. Here,
$\omega_A$ is associated to the curvature of the potential at point $A$,
$\gamma$ is the friction coefficient and KHD stands for Kramers high-damping
limit.

This formula can be used directly to calculate the reaction rate by performing
the integral numerically, and this is what we will do in this section.
However, before doing that, let us introduce two simple approximations that
are instructive and give some insights about the general features of the
problem. For a large barrier, a narrow region around the maximum gives most of
the contribution to the integral. In many cases we can use the so-called
parabolic barrier approximation\cite{Kramers1940}: $U_{\rm near \; B} \simeq
U_B - \frac{1}{2} \omega_B^2 (q-q_B)^2$. Then
\begin{equation}
r_{\rm KHD}^{\rm pb} \simeq
  \frac{\omega_A \omega_B}{2\pi \gamma} e^{-U_B/k_{\rm B}T} \,.
\end{equation}

\begin{figure}[t]
\centering{\includegraphics[angle=-90,width=8.cm]{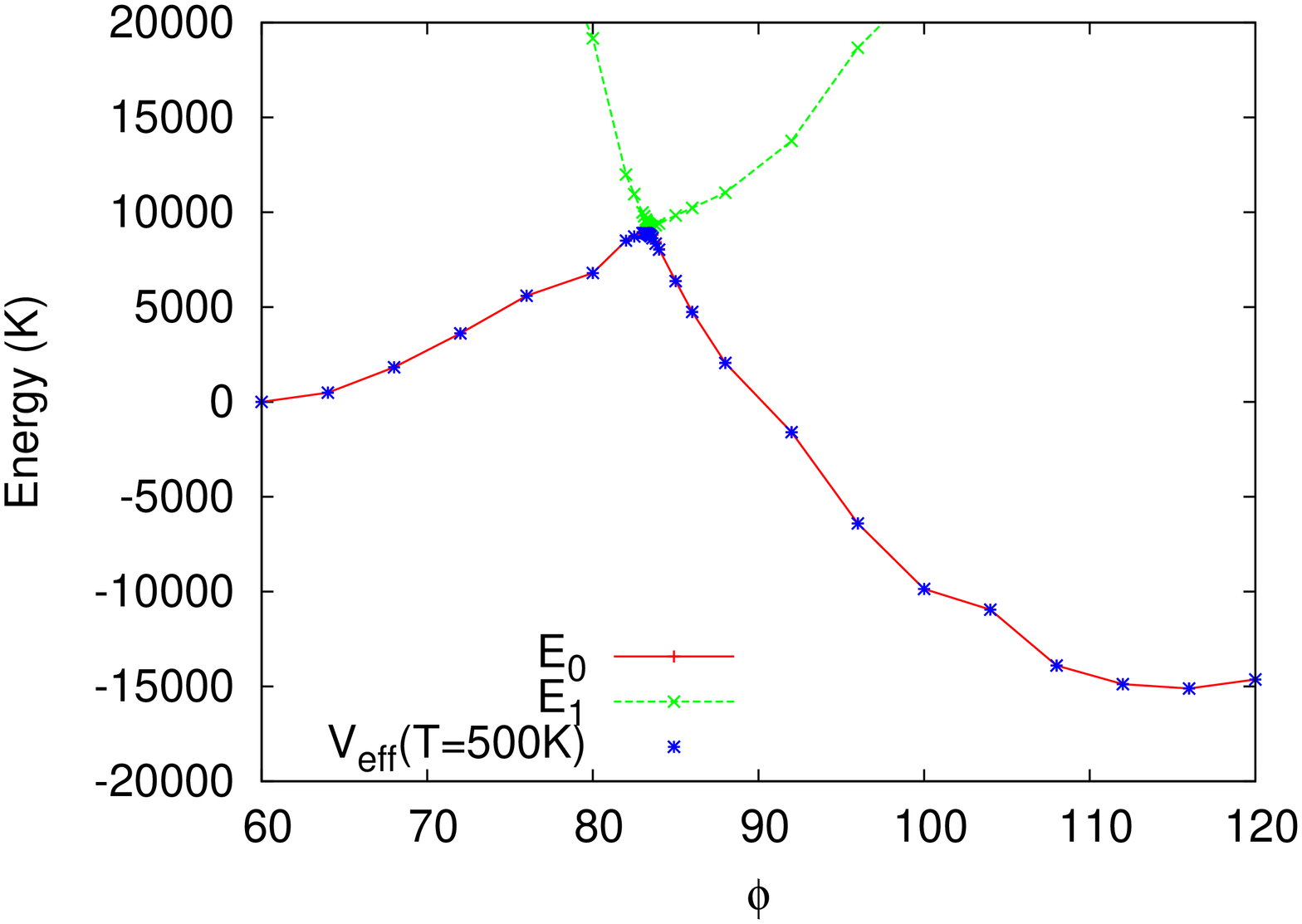}}
\centering{\includegraphics[angle=-90,width=8.cm]{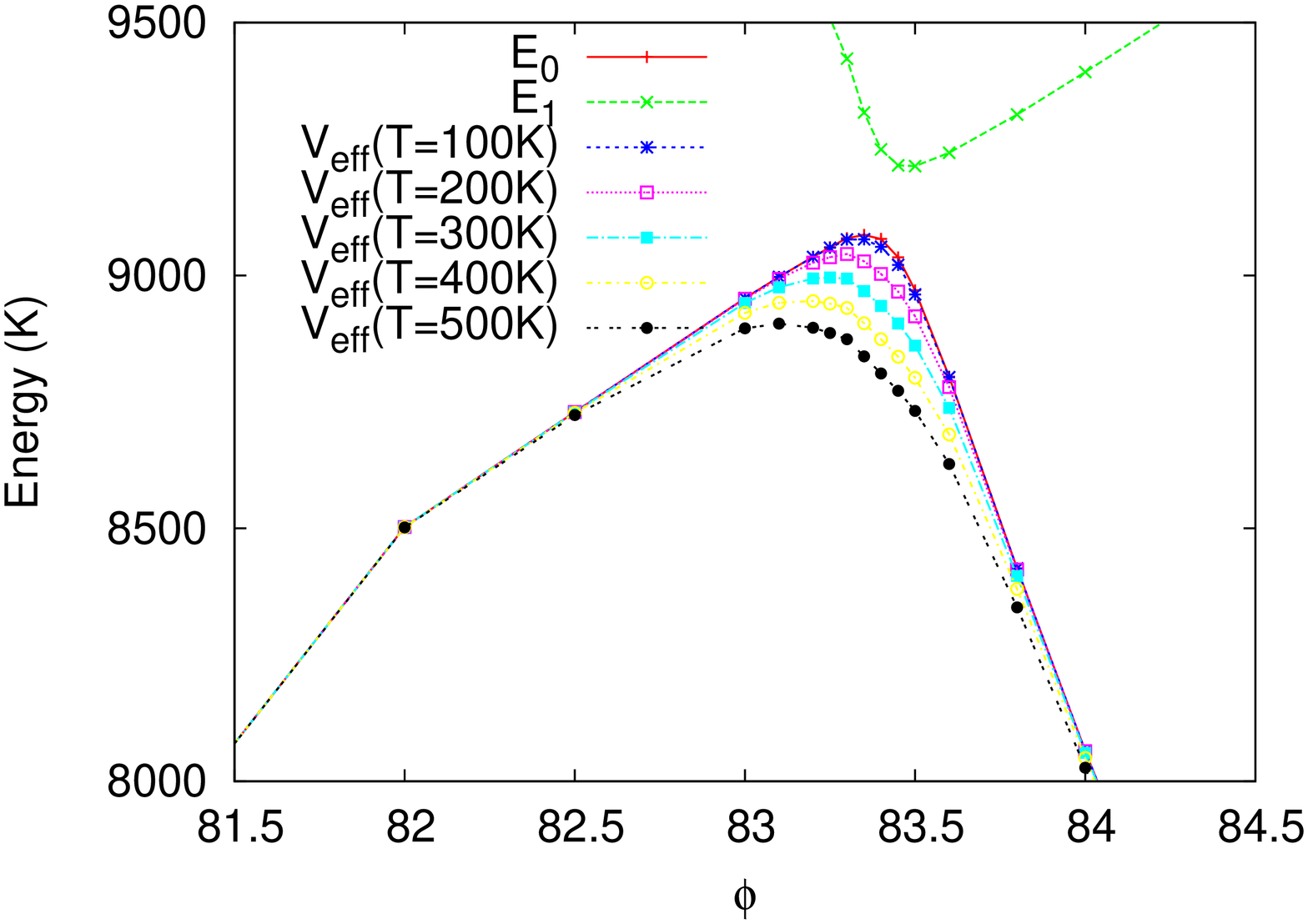}}
\caption{Calculated potential energy profiles for ozone. Top: $E_0(\phi)$, $E_1(\phi)$ and
$V_{\rm eff}(\phi)$ at 500 K. State A lies at $\phi_A = 60^o$ ($E_0(\phi_A)=0$
K) and the barrier is located at approximately $\phi_B = 83.4^o$
($E_0(\phi_B)=9094$ K). Bottom: Potential profiles close to the barrier.
$E_0(\phi)$, $E_1(\phi)$ and $V_{\rm eff}(\phi)$ for $T=$100, 200, 300, 400
and 500 K.}
\label{pp}
\end{figure}

In this situation, we have
\begin{equation}
\label{eq:rate_eff_par}
\frac{r_{\rm eff}}{r_0} \simeq \frac{\omega_B^{\rm eff}(T)}{\omega_B}
 e^{-(U_{\rm eff}(T)-U_0)/k_{\rm B}T} \,.
\end{equation}

Now, if $V_{\rm eff} \sim E_0$ beyond the barrier and $V_{\rm eff} <
E_0$ at the barrier, we expect $\omega_B^{\rm eff} < \omega_B$
(see also the end of Sec.~\ref{sec:results}).
Hence, the prefactor effect opposes the exponential one:
The rate will typically diminish because of the changes in the
prefactor, but it will typically increase due to the changes in the
barrier. In any case, given its exponential dependence, the effect of
the barrier reduction is expected to be dominant in most cases.

Special care must be taken if the region of the potential close to the barrier
(i.e., the one that contributes significantly to the integral in
Eq.~(\ref{I})) cannot be accurately approximated by a parabolic function
around its maximum. Another common approximation to compare with is given
by a cusp barrier:
\begin{equation}
\label{eq:U_cusp}
U_{\rm near \ B} \simeq
\begin{cases}
U_B + m_1 (q - q_B) & \quad \mathrm{if} \quad q \leq q_B \\
U_B - m_2 (q - q_B) & \quad \mathrm{if} \quad q > q_B \\
\end{cases}
\end{equation}
where $m_1,m_2 > 0$ are the slopes at each side of the barrier. In this case,
the activation rate can be written as
\begin{equation}
r_{\rm KHD}^{\rm cb} \simeq \frac{\omega_A}{2\pi \gamma}
  \sqrt{\frac{2\pi}{k_{\rm B}T}} \, \lambda \, e^{-U_B/k_{\rm B}T} \,,
\end{equation}
where $\lambda := m_1 m_2 / (m_1 + m_2)$, and we have
\begin{equation}
\label{eq:rate_eff_cusp}
\frac{r_{\rm eff}}{r_0} \simeq \frac{\lambda^{\rm eff}(T)}{\lambda}
 e^{-(U_{\rm eff}(T)-U_0)/k_{\rm B}T} \,.
\end{equation}

\begin{figure}[t]
\centering{\includegraphics[angle=-90,width=8.cm]{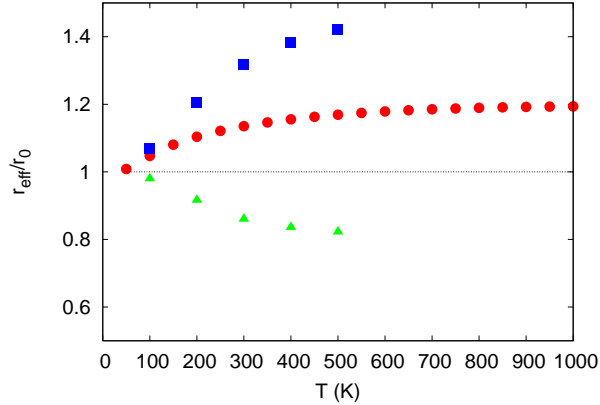}}
\caption{$r_{\rm eff}/r_0$ ratio as a function of $T$ (red circles) as
computed numerically using Eq.~(\ref{I}). We also show the value of the
exponential reduction factor $e^{\delta(T)/k_{\rm B}T}$ (blue squares) and the
prefactor change $b(T)$ estimated from Eq.~(\ref{bT}) (green triangles).}
\label{re_r0}
\end{figure}

In conclusion, we can write in both cases
\begin{equation}
\label{bT}
\frac{r_{\rm eff}}{r_0} \simeq b(T) \ e^{\delta(T)/k_{\rm B}T} \,,
\end{equation}
where $\delta(T) := U_0-U_{\rm eff}(T) > 0$ is the barrier reduction, and
$b(T)$ accounts for the changes in the prefactor. For the simplest case where
$E_0(q)$ and $V_{\rm eff}(q)$ show a maximum at the same coordinate $q_B$ (the
position of the maximum is not affected by the new terms), it is easy to see
that $e^{\delta(T)/k_{\rm B} T}\simeq(1+\ e^{-\Delta /k_{\rm B} T})$ with
$\Delta:= E_1(q_B) - E_0(q_B)$, where we have only considered the first
excited state $E_1$.

As a working case, we will consider now the case of the ring opening reaction
of ozone, which has been previously studied in Ref.~\onlinecite{Alonso2010}.
Fig.~\ref{pp} shows the potential profile for ozone as a function of the
opening angle $\phi$ at different temperatures. Since this is enough for our
purposes, we consider only the gsBO PES, $E_0(\phi)$, and the first excited
state one, $E_1(\phi)$ ---taken from Ref.~\onlinecite{Alonso2010}, where they
where computed using the CASSCF method (complete active-space self-consistent
field).\cite{Roos} The effective potential $V_{\rm eff}(\phi)$ is constructed
with them. As we can clearly see in the bottom graph, the potential profile is
modified by the inclusion of the new term corresponding to $E_1(\phi)$ only in
a narrow region close to the barrier. Also, in this case, the potential
barrier is close to 9000 K and thus $U/k_{\rm B}T \gg 1$ for the temperature
range of interest, which justifies the basic approximations behind the
formulae in this section. However, there exist a clear asymmetry between
$E_0(\phi)$ and $E_1(\phi)$. As a consequence, the barrier location moves with
$T$, and the barrier energy reduction $\delta(T)$ does not follow the expected
$T$ dependence. In addition, as we can see in Fig.~\ref{pp}, neither the
parabolic nor the linear cusp approximations will be suitable to fit the
barrier profile close to the maximum. Hence, we have directly used
Eq.~(\ref{I}) to numerically compute the $r_{\rm eff}/r_0$ ratio of the
system. The results are presented in Fig.~\ref{re_r0}, where we can see that,
in spite that the small barrier reduction observed, the activation rate
changes up to a 20\% upon the inclusion of the excited electronic states. This
indicates that the non-adiabatic effects associated with low-lying states
should be included in any analysis of this kind if one aims for high accuracy
in the predictions. Besides, in the same Fig.~\ref{re_r0}, we also plot the
contributions to the ratio $r_{\rm eff}/r_0$ coming from the changes both in
the potential barrier heigth and shape. As expected the increase of the rate
motivated by the barrier reduction is moderated by the prefactor change. The
two effects work against each other, and the exponential dependence on the
barrier overcomes the influence of the curvature.

\section{Conclusions}
\label{sec:conclusions}

In this work, we have shown that the ground-state PES, as defined in the
Born--Oppenheimer approximation and typically used for many applications in
chemistry, physics and materials science, is not the only electronic state
that significantly contributes to the theoretical prediction and calculation
of thermodynamic observables at non-zero temperature already for small
molecular systems. Although this fact was probably expected by the reader, we
provide actual numerical examples of relevant observables being significantly
modified by the inclusion of thermally-activated low-lying excited electronic
states in real systems at not-too-high temperatures: The average bond length
of the Manganase dimer is shown to change in an amount which is accessible to
modern experimental techniques, while the reaction rate of the ring opening of
ozone is shown to change up to a 20\%. Moreover, our compact effective
potential, which includes any number of such states and which, despite its
simplicity, is able to produce the correct Boltzmann weight for the electronic
subsystem ---a long sought property in the field of quantum-classical models.

Also, as discussed in section Sec.~\ref{sec:methods}, a general
result when using our effective potential is that, in any avoided crossing
situation, the curvature on the top of the barrier is smaller than the gsPES
curvature. As shown in Ref.~\onlinecite{Bothma2010}, if the curvature is
smaller, the tunneling below the energy barrier will occur at lower
temperature. Therefore, the inclusion of low-lying electronic states is
important if one wants to adequately discriminate possible quantum-like
behaviour of the nuclei from simple classical effects due to the direct
influence of temperature on the potential ---for example in biological
systems.\cite{Bothma2010, Iyengar2008}

Additionally, although the effective potential used in this work has been
derived in Ref.~\onlinecite{Alonso2010} assuming classical nuclei and
equilibrated electrons, it could also be used as an external potential to
perform calculations on quantum nuclei. This procedure would allow to study
low temperatures, while still including a possible correction due to
electronic excitations. Notice, however, that, although the use of the gsPES
in the BO approximation as a potential for quantum nuclei can be rigorously
justified (see, e.g., Eq.~(17) in Ref.~\onlinecite{Doltsinis2002}), in the
case of the effective potential used in this work, its use in this manner is
not justified in principle because of its intrinsic non-adiabatic origin.

Finally, it is also reasonable to expect that the use of our temperature
dependent effective potential provides new insights that could lead to answer
the intriguing question of the thermodynamical stability of some diatomic
dications,\cite{Corral2011} an issue we plan to address in the next future.

\section*{Acknowledgements}

We thank A. Rey for insightful comments and illuminating discussions that have
greately contributed to the work we present here.

This work has been supported by the research grants E24/1 and E24/3 (DGA,
Spain), FIS2008-01240, FIS2009-13364-C02-01, FIS2009-12648 and FIS2011-25167
(MICINN, Spain, cofinanced by FEDER funds), and ARAID and Ibercaja grant for
young researchers (Spain). A.R. acknowledges funding from European Research
Council Advanced Grant DYNamo (ERC-2010-AdG - 215 Proposal No. 267374),
Spanish projects (FIS2010-21282-C02- 216 01 and PIB2010US-00652), Grupos
Consolidados UPV/EHU 217 del Gobierno Vasco (IT-319-07), ACI-Promociona
(ACI2009- 218 1036), European Community project THEMA (Contract 219 No.
228539), Ikerbasque, and SGIker ARINA (UPV/EHU).

\bibliography{EffectivePotentialExpansionRates}

\begin{thebibliography}{47}
\bibitem [{ent()}]{entropy}%
  \BibitemOpen
  \href@noop {} {}\bibinfo {note} {\textcolor{blue}{ Let us assume that the
  probability distribution function of a hybrid quantum-classical system can
  indeed be represented by a ``parameterized'' density matrix
  $\hat{\rho}(R,P)$, where the parameters $(R,P)$ are the classical variables.
  Likewise, any observable can be represented by a parameterized operator
  $\hat{O}(R,P)$. The average of any observable can be computed as:
  \begin{equation}\nonumber \langle O \rangle = {\rm Tr}\int{\rm d}R\int{\rm
  d}P\;\hat{O}(R,P)\hat{\rho}(R,P)\,. \end{equation} Then, the
  $\hat{\rho}(R,P)$ that maximizes the entropy, defined as
  \begin{equation}\nonumber S[\hat{\rho}] = -k_{\rm B}\int{\rm d}R{\rm d}P {\rm
  Tr} \left[ \hat{\rho}(R,P){\rm log}\hat{\rho}(R,P) \right]\,, \end{equation}
  subject to the constraints $\int{\rm d}R{\rm d}P {\rm Tr} \hat{\rho}(R,P) =
  1$ and $\langle H \rangle = E$, is precisely given by Eq.~(2.4) in the paper.
  }}\BibitemShut {NoStop}%
\end{thebibliography}

\end{document}